\documentclass[twocolumn,showpacs,superscriptaddress]{revtex4} %
\usepackage{epsfig, amsmath,amssymb,graphicx,graphics}
\usepackage{natbib}
\usepackage[english]{babel}

\begin{document}
\title{
\begin{center}
DYNAMICS OF BOSE-EINSTEIN CONDENSATE\\
WITH ACCOUNT OF PAIR CORRELATIONS
\end{center}
}
\author{Yu.M. Poluektov}
\email{yuripoluektov@kipt.kharkov.ua} %
\affiliation{National Science Center ``Kharkov Institute of Physics and Technology'', %
Akhiezer Institute for Theoretical Physics, 61108 Kharkov, Ukraine} %
\affiliation{ V.N. Karazin National University Sq. Svobody 4, 61077 Kharkov, Ukraine} %
\author{A.M. Arslanaliev} %
\email{arslanaliev.kh@gmail.com} %
\affiliation{ V.N. Karazin National University Sq. Svobody 4, 61077 Kharkov, Ukraine} %

\begin{abstract}
The system of dynamic equations for Bose-Einstein condensate at zero
temperature with account of pair correlations is obtained. The
spectrum of small oscillations of the condensate in a spatially
homogeneous state is explored. It is shown that this spectrum has
two branches: the sound wave branch and the branch with an energy
gap.
\vspace{01mm}\newline%
{\bf Key words}: Bose-Einstein condensate, anomalous and normal
averages, pair correlations, sound branch of elementary excitations,
elementary excitations with energy gap.
\end{abstract}
\pacs{ 67.85.Jk, 67.10.-j } %
\maketitle

\section{Introduction}
Bose-Einstein condensate of a low-density system of weakly
interacting Bose particles at zero temperature is usually described
by the Gross-Pitaevskii equation \cite{Gross61,Pit61}, which is
nowadays widely used for study of the condensates created in
magnetic and laser traps \cite{PitStr03,PitSm02}. The
Gross-Pitaevskii equation is obtained in the self-consistent field
approximation, where the short-range correlations between particles
are neglected. In this case, the Bose system is described in terms
of the coherent  state vector\cite{YuM11}. Meanwhile, the
consideration of the pair correlations, being essential at short
distances, turns out to be important even for systems with low
densities since it leads to some qualitatively new results. For
example, in a dilute classical gas the account for pair correlations
allows us to obtain the integral of collisions in the kinetic
equation and, therefore, all effects which are described by the
Boltzmann equation \cite{Bog70}. \vspace{00mm} %

In the present work, we obtain the system of dynamical equations in
the approximation in which besides the one-particle anomalous
averages, the two-particle correlations are  also taken into account
but the correlations of more number of particles are neglected.
Small oscillations on the background of a spatially homogeneous
equilibrium state are studied. It is shown that when considering
pair correlations in Bose-Einstein condensate, there exist two
branches of elementary excitations. One of them is acoustic, another
one has an
energy gap in the long wavelength limit. %
$\quad\vspace{18mm}$ %

\section{Equations for averages of the field operators}
An arbitrary operator in the Heisenberg representation $A =
e^{i\frac{H}{\hbar}}A(0)e^{-i\frac{H}{\hbar}}$  satisfies the
dynamic equation
\begin{equation} \label{01}
\begin{array}{ll}
\displaystyle{%
  i\hbar\frac{\partial A}{\partial t} = [H, A],%
}%
\end{array}
\end{equation}
where in the second-quantized representation the Hamiltonian can be
written as a sum of operators of the kinetic energy and the energy
of pair interaction $H=H_1+H_2$,
\begin{equation} \label{02}
\begin{array}{ll}
\displaystyle{%
 H_1 = \int d{\bf r}_1d{\bf r}_2 H({\bf r}_1,{\bf r}_2)\Psi^\dagger({\bf r}_1, t)\Psi({\bf r}_2, t),%
}\vspace{2mm}\\ %
\displaystyle{%
 \hspace{35mm} H_2 = %
}\vspace{2mm}\\
\displaystyle{\hspace{-1.4mm}
    =\!\frac{1}{2}\!\int \!\!d{\bf r}_1d{\bf r}_2 U(|{\bf r}_1 - {\bf r}_2|)\Psi^\dagger({\bf r}_1, t)\Psi^\dagger({\bf r}_2, t)\Psi({\bf r}_2, t)\Psi({\bf r}_1, t).
}
\end{array}
\end{equation}
Here
\begin{equation} \label{03}
\begin{array}{ll}
\displaystyle{%
  H({\bf r}_1,{\bf r}_2) = - \frac{\hbar^2}{2m}\Delta_1\delta({\bf r}_1 - {\bf r}_2) + [U_0({\bf r}_1) - \mu]\delta({\bf r}_1 - {\bf r}_2),%
}%
\end{array}
\end{equation}
$m$ is the bose-particle mass, $U_0({\bf r})$ -- the energy of
particle in the external field, $U(|{\bf r}_1 - {\bf r}_2|)$ -- the
particle interaction potential. The field operators satisfy the
usual commutation relations. Let $\langle \Psi\rangle$ is the
average of field operator. Then we can write down our field operator
separating the $c$-number and operator parts:
\begin{equation} \label{04}
\begin{array}{ll}
\displaystyle{%
  \Psi = \langle \Psi\rangle + \xi, \;\;\; \Psi^\dagger = \langle \Psi\rangle^* + \xi^*,%
}%
\end{array}
\end{equation}
The operator part is defined so that it meets the obvious relations:
\begin{equation} \label{05}
\begin{array}{ll}
\displaystyle{%
  \langle\xi\rangle = \langle \xi^\dagger\rangle = 0. \
}%
\end{array}
\end{equation}
Here the averaging is implied in the sense of the quasiaverages for
systems with broken phase symmetry \cite{Bog71, YuM97}. We will
assume that the normal averages, which are invariant under the phase
transformation of the field operators $\Psi \rightarrow \Psi' =
e^{i\alpha}\Psi$, and the anomalous averages, where this invariance
is broken, are both nonzero. It is to be noted that the property of
superfluidity is connected with just the existence of these
anomalous averages.

We introduce the the following notation for the anomalous average of
the field operator\vspace{-0mm}
\begin{equation} \label{06}
\begin{array}{ll}
\displaystyle{%
  \eta({\bf r},t) = \langle\Psi({\bf r},t)\rangle ,\;\; \eta^* = \langle \Psi^\dagger({\bf r},t)\rangle.
}%
\end{array}
\end{equation}
The averages of the products of several field operators can be
written in terms of the averages of the products of operators
$\xi,\xi^\dagger$, which will be called as the overcondensate
operators . For example, for the case of two field operators, using
Eq.(5), we have\vspace{-0mm}
\begin{equation} \label{07}
\begin{array}{ll}
\displaystyle{%
  \langle \Psi^\dagger({\bf r})\Psi({\bf r}')\rangle = \eta^*({\bf r})\eta({\bf r}') + \langle\xi^\dagger({\bf r})\xi({\bf r}')\rangle,
}\vspace{2mm}
\\ %
\displaystyle{%
  \langle \Psi({\bf r})\Psi({\bf r}')\rangle = \eta({\bf r})\eta({\bf r}') + \langle\xi({\bf r})\xi({\bf r}')\rangle,%
}\vspace{2mm}
\\ %
\displaystyle{%
  \langle \Psi^\dagger({\bf r})\Psi^\dagger({\bf r}')\rangle = \eta^*({\bf r})\eta^*({\bf r}') + \langle\xi^\dagger({\bf r})\xi^\dagger({\bf r}')\rangle.%
}
\end{array}
\end{equation}
Similarly, we can write down the averages of a greater number of
field operators. They will also contain the averages of a greater
number of the overcondensate operators of the form %
$\langle \xi^\dagger({\bf r}_1)\xi^\dagger({\bf r}_2)\xi({\bf r}_3)\rangle$, %
$\langle \xi^\dagger({\bf r}_2)\xi^\dagger({\bf r}_3)\xi^\dagger({\bf r}_4)\rangle$, %
$\langle \xi^\dagger({\bf r}_1)\xi^\dagger({\bf r}_2)\xi^\dagger({\bf r}_3)\xi^\dagger({\bf r}_4)\rangle$ %
and etc.
Setting successively the operator $A$ in the Heisenberg equation (1)
equal to $\Psi^\dagger\Psi$, $\Psi\Psi$, $\Psi^\dagger\Psi^\dagger, \ldots\,$ %
and carrying out averaging, we will get the infinite chain of
coupled equations for the averages %
$\langle\Psi\rangle$, $\langle\Psi^\dagger\Psi\rangle$,
$\langle\Psi\Psi\rangle$, $\langle\Psi^\dagger\Psi^\dagger\rangle\ldots\,$, %
which is similar to the Bogoliubov-Born-Green-Kirkwood-Yvon chain
[6] in the kinetic theory of classical gases.

So, the equation for the average of the field operator (6) has the form:
\begin{equation} \label{08}
\begin{array}{ll}
\displaystyle{%
  i\hbar\frac{\partial\eta({\bf r})}{\partial t} = \int H({\bf r},{\bf r}'')\eta({\bf r}'')d{\bf r}'' +
}\vspace{2mm}
\\ %
\displaystyle{%
\hspace{14mm}
   + \int U(|{\bf r} - {\bf r}''|)\langle \Psi^\dagger({\bf r}'')\Psi({\bf r}'')\Psi({\bf r})\rangle d{\bf r}''.
}
\end{array}
\end{equation}
and the equations for the normal and anomalous pair-wise
correlations are written as follows \vspace{-0mm}
\begin{equation} \label{09}\hspace{-0.5mm}
\begin{array}{ll}
\displaystyle{%
  i\hbar\frac{\partial\langle\Psi^\dagger({\bf r})\Psi({\bf r}')\rangle}{\partial t} = %
}\vspace{2mm}
\\ %
\displaystyle{%
  \! =\!\!\int\!\!\Big[ H\!({\bf r}'\!,{\bf r}'')\langle\Psi^\dagger\!({\bf r})\Psi({\bf r}'')\rangle -\! H^*\!({\bf r},{\bf r}'')\langle\Psi^\dagger\!({\bf r}'')\Psi({\bf r}')\rangle\Big] d{\bf r}'' - %
}\vspace{2mm}\\
\displaystyle{%
    - \int\!\Big[ U(|{\bf r}-{\bf r}''|)-U(|{\bf r}'-{\bf r}''|)\Big] \times %
}\vspace{2mm}
\\
\displaystyle{
  \hspace{31mm}\times\langle \Psi^\dagger({\bf r})\Psi^\dagger({\bf r}'') \Psi({\bf r}'')\Psi({\bf r}')\rangle\, d{\bf r}'',
}
\end{array}
\end{equation}
\begin{equation} \label{10}
\begin{array}{ll}
\displaystyle{%
  i\hbar\frac{\partial\langle\Psi({\bf r})\Psi({\bf r}')\rangle}{\partial t} = %
}\vspace{2mm}
\\ %
\displaystyle{%
=\!\int\!\!\Big[H\!({\bf r},{\bf r}'')\langle\Psi({\bf r}')\Psi({\bf r}'')\rangle +\! H\!({\bf r}',{\bf r}'')\langle\Psi({\bf r})\Psi({\bf r}'')\rangle\Big] d{\bf r}''+ %
}\vspace{2mm}%
\\ %
\displaystyle{%
    +\,U({\bf r},{\bf r}')\langle\Psi({\bf r})\Psi({\bf r}')\rangle \,- %
}\vspace{2mm}
\\ %
\displaystyle{%
    - \int\!\Big[ U(|{\bf r}-{\bf r}''|) + U(|{\bf r}'-{\bf r}''|)\Big]\times %
}\vspace{2mm}
\\
\displaystyle{
   \hspace{30mm} \times\langle\Psi^\dagger({\bf r}'')\Psi^\dagger({\bf r}'')\Psi({\bf r})\Psi({\bf r}')\rangle \,d{\bf r}''. %
}
\end{array}
\end{equation}
In the following we will describe the condensate by using the
one-particle averages (6) and restrict ourselves to considering only
pair correlations of the overcondensate operators introduced by
relations (4), having defined the following correlation functions:
\begin{equation} \label{11}
\begin{array}{ll}
\displaystyle{%
 g({\bf r}, {\bf r}',t) = \langle \xi^\dagger({\bf r},t)\xi({\bf r}',t)\rangle,
}\vspace{2mm}
\\
\displaystyle{%
 \tau({\bf r}, {\bf r}',t) = \langle \xi({\bf r},t)\xi({\bf r}',t)\rangle,
}\vspace{2mm}
\\
\displaystyle{%
\tau^*({\bf r}, {\bf r}',t) = \langle \xi^\dagger({\bf r},t)\xi^\dagger({\bf r}',t)\rangle.
}\vspace{2mm}
\\
\end{array}
\end{equation}
The averages of a greater number of the overcondensate operators
will be neglected that seems to be acceptable for sufficiently
diluted systems. Functions (11) have the obvious symmetry properties
\begin{equation} \label{12}
\begin{array}{ll}
\displaystyle{%
g({\bf r}, {\bf r}',t) = g^*({\bf r}', {\bf r},t),\;\tau({\bf r}, {\bf r}',t) = \tau({\bf r}', {\bf r},t),
}\vspace{2mm}
\\
\displaystyle{%
 \hspace{20mm}\tau^*({\bf r}, {\bf r}',t) = \tau^*({\bf r}', {\bf r},t).
}
\end{array}
\end{equation}
When only the pair correlations are taken into account, from
(8)\,--\,(10) one can get the closed system of equations for
functions $\eta({\bf r},t)$ $g({\bf r},{\bf r}',t)$ $\tau({\bf
r},{\bf r}',t)$
\begin{equation} \label{13}
\begin{array}{ll}
\displaystyle{%
  i\hbar\frac{\partial\eta({\bf r})}{\partial t} = -\frac{\hbar^2}{2m}\Delta\eta({\bf r}) + [U_0({\bf r}) - \mu]\eta({\bf r}) +
}\vspace{2mm}
\\ %
\displaystyle{%
+ \int U(|{\bf r} - {\bf r}''|)\Big[|\eta({\bf r}'')|^2\eta({\bf r})
+ \eta^*({\bf r}'')\tau({\bf r},{\bf r}'') +  }
\vspace{2mm}\\
\hspace{15mm}   + \eta({\bf r}'')g^*({\bf r},{\bf r}'') + \eta({\bf r})g({\bf r}'',{\bf r}'')\Big], %
\end{array}
\end{equation}
\vspace{-4mm}
\begin{equation} \label{14}
\begin{array}{ll}
\displaystyle{%
  i\hbar\frac{\partial\tau({\bf r},{\bf r}')}{\partial t} = U(|{\bf r}-{\bf r}'|)\eta({\bf r})\eta({\bf r}') + U(|{\bf r}-{\bf r}'|)\tau({\bf r},{\bf r}') -
}\vspace{2mm}
\\ %
\displaystyle{%
    -\frac{\hbar^2}{2m}(\Delta + \Delta')\tau({\bf r},{\bf r}') + [U_0({\bf r}) + U_0({\bf r}') - 2\mu]\tau({\bf r},{\bf r}')+ %
}\vspace{2mm}
\\ %
\displaystyle{%
    +\! \int\!\! d{\bf r}''U(|{\bf r} - {\bf r}''|)\Big[|\eta({\bf r}'')|^2\tau({\bf r},{\bf r}') + \eta({\bf r})\eta^*({\bf r}'')\tau({\bf r}',{\bf r}'') +
    }\vspace{2mm}
\\
\displaystyle{
     +\, \eta({\bf r})\eta({\bf r}'')g({\bf r}'',{\bf r}')\Big] \! + \! \int\!\! d{\bf r}'' U(|{\bf r}'-{\bf r}''|)\Big[|\eta({\bf r}'')|^2\tau({\bf r},{\bf r}') +
}\vspace{2mm}
\\ %
\displaystyle{%
\hspace{14mm}    +\, \eta({\bf r}')\eta^*({\bf r}'')\tau({\bf r},{\bf r}'') + \eta({\bf r}')\eta({\bf r}'')g({\bf r}'',{\bf r})\Big], %
}
\end{array}
\end{equation}
\vspace{-5mm}
\begin{equation} \label{15}
\begin{array}{ll}
\displaystyle{%
  i\hbar\frac{\partial g({\bf r},{\bf r}')}{\partial t}\! =\!\frac{\hbar^2}{2m}(\Delta - \!\Delta')g({\bf r},{\bf r}')-\! [U_0({\bf r})\! -\! U_0({\bf r}'\!)]g({\bf r},{\bf r}') - %
}\vspace{2mm}
\\ %
 \displaystyle{%
     -\int \!d{\bf r}''U(|{\bf r} - {\bf r}''|)\Big[|\eta({\bf r}'')|^2g({\bf r},{\bf r}') + \eta^*({\bf r})\eta({\bf r}'')g({\bf r}'',{\bf r}') +
}\vspace{2mm}
\\ %
\displaystyle{
    + \eta^*\!({\bf r})\eta^*\!({\bf r}'')\tau({\bf r}'',{\bf r}')\Big] \! + \!\! \int \!\! d{\bf r}''U(|{\bf r}'-{\bf r}''|)\Big[|\eta({\bf r}'')|^2g({\bf r},{\bf r}')+%
}\vspace{2mm}
\\
\displaystyle{%
\hspace{14mm} + \eta({\bf r}')\eta^*({\bf r}'')g({\bf r},{\bf r}'') + \eta({\bf r}')\eta({\bf r}'')\tau^*({\bf r}'',{\bf r})\Big]. %
}
\end{array}
\end{equation}
One should note that this system of equations is invariant under
time-reversal transformation, because along with solutions
$\eta({\bf r},t)$, $g({\bf r},{\bf r}',t)$, $\tau({\bf r},{\bf r}',t)$ %
it also has the solutions
$\eta^*({\bf r},-t)$, $g^*({\bf r},{\bf r}',-t)$, $\tau^*({\bf r},{\bf r}',-t)$. %
Neglecting the pair correlations $\tau({\bf r},{\bf r}',t)$ and
$g({\bf r},{\bf r}',t)$, equation (13) takes the form of the
Gross-Pitaevskii equation [1,2].
In what follows, where it will not cause confusion, as in equations
(13)\,--\,(15), for brevity the explicit time-dependance of the
averages will be omitted.

The average of the operator of the total number of particles $N$ is
given by the formula
\begin{equation}\label{16}
\langle N\rangle = \int d{\bf r}[\eta^*({\bf r},t)\eta({\bf r},t) + g({\bf r},{\bf r},t)], %
\end{equation}
and the particle number density is, obviously, $n({\bf r},t) =
\eta^*({\bf r},t)\eta({\bf r},t) + g({\bf r},{\bf r},t)$.

\section{Local form of equations}
Equations (13)\,--\,(15) are integro-differential. While studying
the states that slowly vary on the scales comparable to
characteristic radius $r_0$ of action of the interparticle
interaction potential $U(|{\bf r}-{\bf r'}|)$, we can pass to
differential equations. The pair correlation functions (11) depend
on two coordinates ${\bf r}, {\bf r}'$. It is convenient to
introduce  new variables $\vec{\rho} = {\bf r} - {\bf r}'$, ${\bf R}
= \frac{1}{2}({\bf r} + {\bf r}')$, then \vspace{-0mm}
\begin{equation}\label{17}
\begin{array}{ll}
 \displaystyle{
 \tau({\bf r}, {\bf r}') = \tau\big({\bf R} + \frac{\vec{\rho}}{2}, {\bf R} - \frac{\vec{\rho}}{2}\big) \equiv \tau({\bf R}, \vec{\rho}), %
}
\vspace{2mm}\\
\displaystyle{
g({\bf r}, {\bf r}') = g\big({\bf R} + \frac{\vec{\rho}}{2}, {\bf R} - \frac{\vec{\rho}}{2}\big) \equiv g({\bf R},\vec{\rho}).%
}
\end{array}
\end{equation}
These functions slowly change depending on the pair's center-mass
coordinate ${\bf R}$ on distances of the order of the interparticle
potential radius $r_0$. The correlation functions can be presented
in the form:
\begin{equation}\label{18}
\displaystyle{
    \tau({\bf R},\vec{\rho}) = \sum_{\bf k}\tau_{{\bf k}}({\bf R})e^{i{\bf k}\vec{\rho}},\;\;g({\bf R}, \rho)= \sum_{\bf k}g_{{\bf k}}({\bf R})e^{i{\bf k}\vec{\rho}}.%
}
\end{equation}
In the following, only the term with ${\bf k} = 0$ will be taken
into account in these sums. It means that instead of the exact
functions $\tau({\bf R}, \rho)$ $g({\bf R}, \rho)$, we will use the
functions which are averaged over a macroscopic volume $V_0 \sim
L^3$, where $L \gg r_0$:
\begin{equation*}
\displaystyle{
\tau_0({\bf R})\approx V_0^{-1}\!\int\! \tau({\bf R},\vec{\rho})\,d\vec{\rho},\;\; g_0({\bf R})\approx V_0^{-1}\!\int\! g({\bf R},\vec{\rho})\,d\vec{\rho} .%
}
\end{equation*}
This approximation is acceptable if one considers perturbations on
spatial scales that significantly exceed the radius of action of the
interparticle potential.

It is worth noting that in obtained equations the behavior of the
interparticle interaction potential at short distances plays an
important role. The form of the potential is  poorly known here.
Moreover, for many model potentials such as, for example the
Lennard-Jones potential, it is assumed that at short distances it
goes to infinity. Note also that the use of model potentials which
go to infinity at short distances in some cases leads to
considerable difficulties, because such potentials do not have the
Fourier representation. Meanwhile, the requirement of
``impermeability'' of atoms at arbitrary high pressures is very
strict, since there should exist a pressure at which an atom will be
``crushed'' and stop existing as a separate structural unit.
Therefore, in our opinion, it is physically reasonable and natural
to use the potentials, which take a finite value at short distances.
It should also be noted that the quantum-chemical calculations give
potentials with a finite, albeit large, value at zero \cite{Aziz91,
And93}.

In this local approximation, the system of equations (13)\,--\,(15)
takes the form:
\begin{equation}\label{19}
\begin{array}{ll}
\displaystyle{
i\hbar\frac{\partial \eta ({\bf r})}{\partial t} = -\frac{\hbar^2}{2m}\Delta \eta({\bf r}) + [U_0({\bf r}) - \mu]\eta({\bf r}) + %
}\vspace{2mm}\\
\displaystyle{
+ U_0\left[|\eta({\bf r})|^2\eta({\bf r}) + \eta^*({\bf r})\tau({\bf r}) + 2\eta({\bf r})g({\bf r})\right],
}
\end{array}
\end{equation}
\vspace{-2mm}
\begin{equation}\label{20}
\begin{array}{ll}
\displaystyle{i\hbar\frac{\partial\tau ({\bf r})}{\partial t} = -\frac{\hbar^2}{4m}\Delta \tau({\bf r}) + U(0)\eta^2({\bf r}) + U(0)\tau({\bf r})\,+ %
}\vspace{2mm}\\%
\displaystyle{
 +\, [2U_0({\bf r}) - 2\mu]\tau({\bf r}) + U_0\!\left[4|\eta({\bf r})|^2\tau({\bf r}) + 2\eta^2({\bf r})g({\bf r})\right], %
}
\end{array}
\end{equation}
\begin{equation}\label{21}
\begin{array}{ll}
\displaystyle{
i\hbar\frac{\partial g({\bf r})}{\partial t} = -U_0\left[\eta^{*2}({\bf r})\tau({\bf r}) - \eta^2({\bf r})\tau^*(\bf{r})\right]. %
}\vspace{2mm}
\end{array}
\end{equation}
Here we used the designations: $\tau({\bf R})\equiv \tau({\bf r})$,
$g_0({\bf R})\equiv g({\bf r})$, $U_0\equiv \int U({\bf r})d{\bf
r}$. Since the magnitude of the potential energy of interaction of
atoms at short distances $U(0)$ is poorly known, it will be regarded
as a phenomenological adjustable parameter. For specific
calculations we will use a simple model potential of
``semi-transparent sphere'' form:
\begin{equation}\label{22} %
\displaystyle{ %
U({\bf r}) = \begin{cases}
u,\;\;\;\;\;\;\; r<r_0\,,\\
0,\;\;\;\;\;\;\; r>r_0\,.
\end{cases}
}
\end{equation}
The parameter $u$ is assumed to be positive. In this case $U(0) =
u$, $U_0 = uv$, where $v \equiv \frac{4\pi}{3}r_0^3 $ is the "atomic
volume". The potential (22) has earlier been used in the studies of
the Bose systems (see e.g.~\cite{Brak64}). Further, we will analyze
the system of equations (19)\,--\,(21) and use the potential (22)
for some estimations.

\section{Equilibrium spatially homogeneous state}
In this section we will consider the equilibrium state of a
spatially homogeneous system. In this case, the functions $\eta({\bf
r})\equiv \eta$, $\tau({\bf r})\equiv \tau$, $g({\bf r})\equiv g$ do
not depend on coordinates and time. For sufficiently weak
interaction, the most of particles will be in the single-particle
condensate at zero temperature ~\cite{Bog47, Bog84}. Therefore, we
will assume the normal correlation function $g = 0$ at the
equilibrium state, so that the equilibrium particle number density
$n = |\eta|^2$. Then, in the absence of the external field, from
(19)\,--\,(21) there follow the equations which determine the
equilibrium state of the system:
\begin{equation}\label{23}
\displaystyle{
   -\mu\eta + U_0\left[|\eta|^2\eta + \eta^*\tau\right] = 0,
}
\end{equation}
\begin{equation}\label{24}
\displaystyle{
    U(0)\eta^2 + \left[U(0) - 2\mu + 4U_0|\eta|^2\right]\tau = 0,
}
\end{equation}
\begin{equation}\label{25} %
\displaystyle{
\eta^{*2}\tau - \eta^2\tau^* = 0.
}
\end{equation}
Let's write down the complex quantities extracting their moduli and
phases: $\eta = \eta_0e^{i\alpha}$, $\tau = \tau_0e^{i\beta}$. From
(25) it follows that $\sin{(2\alpha - \beta)} = 0$. Thus, there are
two possibilities $2\alpha - \beta = 0$ or $2\alpha - \beta = \pi$.
We have to choose the second possibility, since only in this case
equations (23),(24) have the physically correct solutions:
\begin{equation} \label{26} %
\displaystyle{
  \eta_0\left[\mu - U_0(\eta_0^2 - \tau_0)\right] = 0,
}
\end{equation}
\begin{equation} \label{27} %
\displaystyle{
    U(0)\eta_0^2 - \left[U(0) + 4U_0\eta_0^2 - 2\mu\right]\tau_0 =0.
}
\end{equation}
With this choice of phase $\tau = -\tau_0e^{2i\alpha}$. After
eliminating the chemical potential $\mu$ from these equations, we get %
\begin{equation} \label{28} %
\displaystyle{ \left[U(0)-2U_0\tau_0\right]\eta_0^2 - [U(0) +
2U_0\tau_0]\tau_0 = 0. }
\end{equation}
Note that there does not exist the solution for the system of
equations (26) and (27) such that only the single-particle
condensate exists $\eta_0 \neq 0$  and the pair condensate is absent
$\tau_0 = 0$. This feature of the Bose systems has been earlier
pointed out~\cite{YuM07}. From (28) it follows the relation between
density and the pair correlation:
\begin{equation} \label{29} %
\displaystyle{
    n \equiv \eta_0^2 = \frac{1 + 2v\tau_0}{1-2v\tau_0}\tau_0.
    }
\end{equation}
Here we have used the relation $U_0 = uv$ ($v$ is the ``atomic
volume''), which was obtained above for the potential (22). Hence it
follows the restriction $v\tau_0 < 1/2$. The chemical potential
\begin{equation} \label{30} %
\displaystyle{ \mu = 4uv\frac{v\tau_0^2}{1-2v\tau_0} }
\end{equation}
proves to be positive. The dependence of the pair correlation
on the particle density is given by the relation
\begin{equation} \label{31} %
\displaystyle{
\tau_0v = \frac{1}{2}\left[\sqrt{\left(nv + \frac{1}{2}\right)^2 + 2nv } - \left(nv + \frac{1}{2}\right)\right]. %
}
\end{equation}
Here $nv = v/\Omega$, where $\Omega$ is the volume per one particle.
In a dilute system $\Omega \gg v$ and $nv \ll1$. In this case $n \approx \tau_0$. %
\begin{figure}[h!]
\vspace{-00mm} \hspace{-01mm}
\includegraphics[width = 0.94\columnwidth]{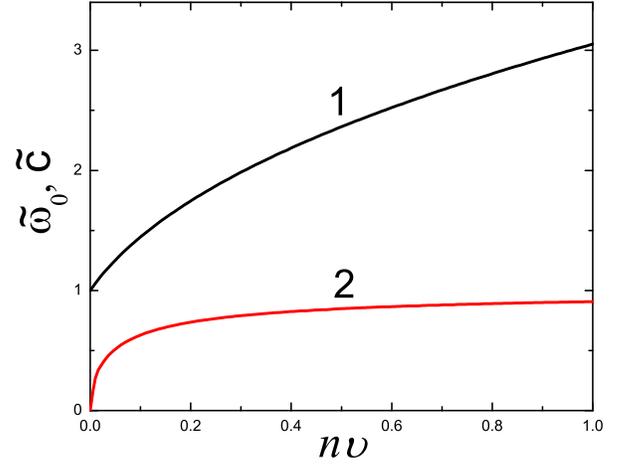} %
\vspace{-04mm}
\caption{\label{Fig01} %
Dependencies of the homogeneous oscillations frequency
$\tilde{\omega}_0 = \omega_0/u$ (curve 1) and the speed of sound
$\tilde{c} = c/c_B$ (curve 2) on density, $c_B = \sqrt{uvn/m}$. %
}%
\end{figure}

\section{SPECTRUM OF Elementary excitations}\vspace{-2mm}
In this section we will consider the propagation of small
perturbations in a spatially homogeneous system. Setting
\begin{equation} \label{32} %
\displaystyle{
    \eta({\bf r}, t)= \eta_0 + \delta\eta({\bf r}, t), \; \tau({\bf r}, t) = \tau_0 + \delta\tau,\; g({\bf r},t)\! = \delta g({\bf r}, t), %
}
\end{equation}
we get from (19)\,--\,(21) the system of linearized equations:
\begin{equation} \label{33} %
\begin{array}{ll}
\displaystyle{ i\hbar\frac{\partial \delta\eta}{\partial t} = -
\frac{\hbar^2}{2m}\Delta\delta\eta + U_0\Big[\left(\eta_0^2 +
\tau_0\right)\delta\eta + (\eta_0^2 - \tau_0)\delta\eta^* +
}\vspace{2mm}
\\
\displaystyle{
 \hspace{35mm}+ \,\eta_0\delta\tau + 2\eta_0\delta g\Big],
}
\end{array}
\end{equation}
\begin{equation} \label{34} %
\begin{array}{ll}
\displaystyle{
i\hbar\frac{\partial \delta\tau}{\partial t} = - \frac{\hbar^2}{4m}\Delta\delta\tau + \Big[U(0) + 2U_0(\eta_0^2 + \tau_0)\Big]\delta\tau +
}\vspace{2mm}
\\
\displaystyle{
 + \,2\Big[U(0)-2U_0\tau_0\Big]\eta_0\delta\eta - 4U_0\eta_0\tau_0\delta\eta^* + 2U_0\eta_0^2\delta g,
}
\end{array}
\end{equation}
\begin{equation} \label{35} %
\begin{array}{ll}
\displaystyle{
i\hbar\frac{\partial \delta g}{\partial t} = - U_0\Big[\eta_0^2(\delta \tau - \delta\tau^*) + 2\eta_0\tau_0(\delta\eta - \delta\eta^*)\Big].
}
\end{array}
\end{equation}
It is convenient to pass from the complex variables $\delta\eta({\bf
r},t)$, $\delta\tau({\bf r}, t)$, $\delta g({\bf r}, t)$ to the real ones %
\begin{equation} \label{36} %
\begin{array}{ll}
\displaystyle{
    \delta\Psi({\bf r},t) = \delta\eta({\bf r},t) + \delta\eta^*({\bf r},t),
}\vspace{2mm}
\\
\displaystyle{
\delta\Phi({\bf r},t) = i\big[\delta\eta({\bf r},t) - \delta\eta^*({\bf r},t)\big],
}\vspace{2mm}
\\
\displaystyle{
   \delta\Theta({\bf r},t) = \delta\tau({\bf r},t) + \delta\tau^*({\bf r},t),
}\vspace{2mm}
\\
\displaystyle{
\delta\Lambda({\bf r},t) = i\big[\delta\tau({\bf r},t) - \delta\tau^*({\bf r},t)\big].
}
\end{array}
\end{equation}
For real quantities the system of linearized equations takes the form:
\begin{equation} \label{37} %
\displaystyle{ \hbar\frac{\partial\delta\Psi}{\partial t} =
\frac{\hbar^2}{2m}\Delta\delta\Phi - U_0\big[2\tau_0\delta\Phi +
\eta_0\delta\Lambda\big], } \displaystyle
\end{equation}
\begin{equation} \label{38} %
\displaystyle{ \hbar\frac{\partial\delta\Phi}{\partial t} =
-\frac{\hbar^2}{2m}\Delta\delta\Psi + U_0\big[2\eta^2_0\delta\Psi +
\eta_0\delta\Theta + 4\eta_0\delta g\big], }
\end{equation}
\begin{equation} \label{39} %
\begin{array}{ll}
\displaystyle{
    \hbar\frac{\partial\delta \Lambda}{\partial t} = -\frac{\hbar^2}{4m}\Delta\delta\Theta + \big[U(0) + 2U_0(\eta_0^2 + \tau_0)\big]\delta\Theta \,+
}\vspace{2mm}
\\
\displaystyle{
    \hspace{15mm}+2\,\big[U(0) - 4U_0\tau_0\big]\eta_0\delta\Psi + 4U_0\eta_0^2\delta g,
}
\end{array}
\end{equation}
\begin{equation} \label{40} %
\begin{array}{ll}
\displaystyle{ \hbar\frac{\partial \delta\Theta}{\partial t} =
\frac{\hbar^2}{4m}\Delta\delta\Lambda - \big[U(0) + \,2U_0(\eta_0^2
+ \tau_0)\big]\delta\Lambda - }\vspace{2mm}
\\
\displaystyle{
    \hspace{25mm}- 2U(0)\eta_0\delta\Phi,
}
\end{array}
\end{equation}
\vspace{-2mm}
\begin{equation} \label{41} %
\displaystyle{ \hbar\frac{\partial \delta g}{\partial t} = U_0\eta_0\big[2\tau_0\delta\Phi + \eta_0\delta\Lambda\big]. } %
\end{equation}
Assuming that the dependence of the fluctuations on the coordinates
and time is of the form $\exp{[i{\bf kr} - \omega t]}$, one gets the
system of homogeneous linear algebraic equations.
From the condition of equality to zero of its determinant we obtain
the biquadratic equation that determines the dispersion laws of
possible excitations:
\begin{equation}\label{42} %
\displaystyle{ (\hbar\omega)^4 - A(\hbar\omega)^2 + B = 0. }
\end{equation}
Here
\begin{equation}\label{43} %
\begin{array}{ll}
\displaystyle{
  A = (\hbar\omega_0)^2  + a_1\varepsilon_k + \frac{5}{4}\varepsilon_k^2,
}\vspace{2mm}
\\
\displaystyle{
    B = b_1\varepsilon_k + b_2\varepsilon_k^2 + b_3\varepsilon_k^3 +\frac{\varepsilon^4_k}{4},
}
\end{array}
\end{equation}
where $\displaystyle{\varepsilon_k = \frac{\hbar^2k^2}{2m}}$ is the
free particle energy. The coefficients in Eq.(43) have the form
\begin{equation} \label{44} %
\begin{array}{ll}
\displaystyle{
    a_1 = U(0) + 4U_0(\eta_0^2 + \tau_0),
}\vspace{2mm}
\\
\displaystyle{
    b_1 = 4U_0^2\eta_0^2\big[U(0)(3\eta_0^2 + 2\tau_0) + 7U_0\tau_0^2\big],
}\vspace{2mm}
\\
\displaystyle{ b_2 = U^2\!(0) \!+ \! 5U_0U\!(0)(\eta_0^2\! + \!
\tau_0) + U_0^2(6\tau_0^2 + 4\eta_0^4 + 17\tau_0\eta_0^2),
}\vspace{2mm}
\\
\displaystyle{
b_3 = U(0) + \frac{5}{2}U_0(\eta_0^2 + \tau_0).
}
\end{array}
\end{equation}
The system admits a solution in the form of spatially homogeneous
oscillations with a frequency $\omega_0$, which is determined by the
following formula:
\begin{equation} \label{45} %
\displaystyle{ (\hbar\omega_0)^2 = U^2(0) + 6U_0U(0)(\eta_0^2 + \tau_0) + 8U_0^2\tau_0^2. } %
\end{equation}
The dependence of this frequency on the density, determined with
account of the relation (31), is shown in Fig.1 (curve 1).

The biquadratic equation (42) has two solutions, which determine two excitation branches:
\begin{equation} \label{46} %
\displaystyle{ (\hbar\omega_{\pm})^2 = \frac{1}{2}\big[A \pm \sqrt{A^2 - 4B}\big]. } %
\end{equation}
The solution $\omega_-$ at small wave numbers gives the sound branch
$\omega^2_- = c^2k^2$, where the square of speed of sound  is
determined by the formula:
\begin{equation} \label{47} %
\displaystyle{
c^2 = \frac{2U_0^2\eta_0^2\big[U(0)(3\eta_0^2 + \tau_0) + 7U_0\tau_0^2\big]}{m\big[U^2(0) + 6U_0U(0)(\eta_0^2 + \tau_0) + 8U_0^2\tau_0^2\big]} . %
}
\end{equation}
The dependence of the speed of sound on density is shown in Fig. 1 (curve 2).

The solution $\omega_+$ corresponds to the excitation branch with an
energy gap. For small $k$ the dependence of the frequency on the
wave number has the form
\begin{equation} \label{48} %
\displaystyle{ \omega_+^2 = \omega_0^2 + \alpha\frac{k^2}{2m}, } %
\end{equation}
where \vspace{2mm}
\begin{equation}
\begin{array}{ll}\label{49} %
\displaystyle{
\alpha = \omega_0^{-2}\big[ U^3\!(0) + 10\,U^2\!(0)U_0(\eta_0^2 + \tau_0)\,  +  %
}\vspace{2mm}\\
\displaystyle{ \hspace{4mm} +\, 4\,U\!(0)U_0^2(8\tau_0^2 + 3\eta_0^4 + 10\eta_0^2\tau_0) + 4U_0^3\tau_0^2(\eta_0^2 + 8\tau_0) \big]. } %
\vspace{2mm}
\end{array}
\end{equation}

Although, strictly speaking, these equations are applicable for the
long-wave excitations, but the solutions (46) give the reasonable
values also for large $k$:
\begin{equation}\label{50} %
\displaystyle{
    \hbar\omega_+ = \varepsilon_k,\;\;\; \hbar\omega_- = \frac{\varepsilon_k}{2}. %
}
\end{equation}
One branch turns into to the dispersion law of a free particle and
the other branch gives the dispersion law of a pair of coupled
particles. The branches of elementary excitations in the Bose system
with account of pair correlations are shown in Fig.2. This figure
also shows the Bogolyubov dispersion law $\hbar\omega_B =
\varepsilon_k(\varepsilon_k + 2uvn) $.

\begin{figure}[h!]
\vspace{1mm}  %
\centering{\hspace{-1mm}\includegraphics[width =1.00\columnwidth]{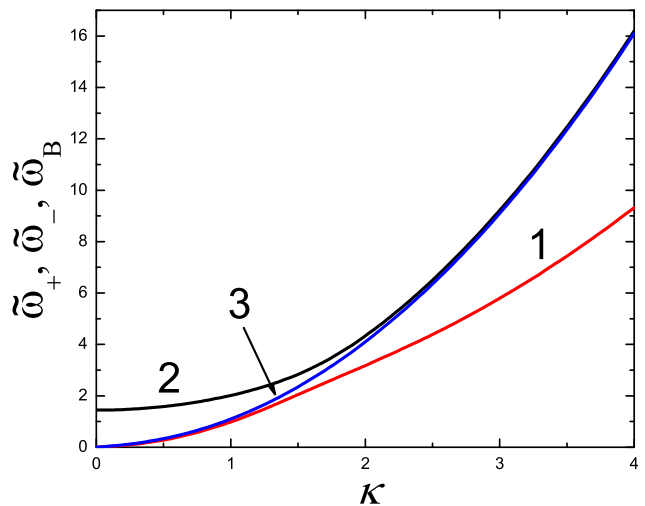}} %
\vspace{-4mm}
\caption{\label{Fig02} %
The dispersion law of the elementary excitations with account of
pairing correlations: 1) sound branch $\tilde{\omega}_- =\omega_-/u$; %
2) branch with an energy gap $\tilde{\omega}_+ = \omega_+/u$ 3) the
Bogolyubov dispersion law $\tilde{\omega}_B = \omega_B/u$, %
$\kappa = k/\sqrt{2mU(0)}$. The calculation is performed for $nv = 0.1$. %
}%
\end{figure}
\vspace{-5mm}

\section{Conclusion}
We have obtained the system of differential equations (19)\,--\,(21) %
which describes the dynamics of the Bose-Einstein condensate with
account of pair correlations. The spectrum of small oscillations in
a spatially homogeneous system was studied. It was shown that there
are two branches of elementary excitations: one branch with the
sound dispersion law in the long wavelength limit and the second
branch which has in this limit an energy gap.

It should be noted that the question of the possible existence in
the Bose systems of excitations with an energy gap has a long
history and has been discussed in many papers (see for e.g.
\cite{Bijl40}--\cite{YuM02}). The possible existence, in addition to
the phonon branch of the spectrum, of the other branch with an
energy gap was discussed at a qualitative level in [13, p. 322].

In the experimental paper \cite{Rubalko} the absorption of microwave
radiation was discovered in the superfluid helium at a frequency of
about 180 GHz, which authors attributed to the creation of single
rotons. However, due to the fact that the momentum of a roton is by
orders of magnitude greater than the momentum of a photon at the
given frequency, it was suggested in \cite{YuM14} that the observed
absorption  owes its existence to the presence of excitations with
energy gap in the superfluid helium. The branches of elementary
excitations (one of which is sound and another has an energy gap),
which are obtained theoretically in this work, can be considered as
a confirmation of the qualitative arguments in favor of the possible
existence of the gap excitations in the superfluid helium and the
need for modification of the energy spectrum of He-II, which were
formulated in \cite{YuM14}.


\vspace{-0mm}
\begin{thebibliography}{99}
\bibitem{Gross61}
  E.P.\,Gross, Structure of a quantized vortex in boson systems, Il Nuovo Cimento \textbf{20}, 454-477 (1961). %
\bibitem{Pit61}
  L.\,Pitaevskii, Vortex lines in an imperfect Bose gas, Sov. Phys. JETP. \textbf{13}, 451-454 (1961). %
\bibitem{PitStr03}
  L.\,Pitaevskii, S.\,Stringari, Bose-Einstein condensation, Oxford  University Press, USA, 492 p. (2003). %
\bibitem{PitSm02}
  C.H.\,Pethick, H.\,Smith, Bose-Einstein condensation in dilute gases, Cambridge University Press, 402 p. (2002). %
\bibitem{YuM11}
 Yu.M.\,Poluektov, The polarization properties of an atomic gas in a coherent state, Low Temp. Phys. \textbf{37}, N12, 986 p. (2011). %
\bibitem{Bog70}
    N.N.\,Bogolyubov, Problems of dynamic theory in statistical physics, in book: Selected works in three volumes, Vol. \textbf{2}, Naukova Dumka, Kiev, 522 p. (1970). %
\bibitem{Bog71}
    N.N.\,Bogolyubov, Quasiaverages in problems of statistical mechanic, in book: Selected works in three volumes, Vol. \textbf{3}, Naukova Dumka, Kiev, 488 p. (1971). %
\bibitem{YuM97}
    Yu.M.\,Poluektov, On self-consistent determination of the quasi-average in statistical physics, Low Temp. Phys. \textbf{23}, N9, 685 p. (1997). %
\bibitem{Aziz91}
R.A.\,Aziz, M.J.\,Slaman, An examination of ab initio result for the helium potential energy curve, J. Chem. Phys. \textbf{94}, 8047\,p. (1991). %
\bibitem{And93}
J.B.\,Anderson, C.A.\,Traynor, B.M.\,Boghosian,  An exact quantum Monte Carlo calculation of the helium-helium intermolecular potential, J. Chem. Phys. \textbf{99}, 345\,p. (1993). %
\bibitem{Brak64}
K. A. Brueckner, Theory of nuclear structure. The many body problem. London, Methuen a. o. (1959). %
\bibitem{Bog47}
N.N.\,Bogolyubov, On the theory of superfluidity, J. Phys. USSR \textbf{11}, 23-32 (1946); Izv. AN SSSR, Ser. Fiz. \textbf{11}(1), 77-90 (1947). %
\bibitem{Bog84}
N.N.\,Bogolyubov and N.N.\,Bogolyubov, Jr., Introduction to quantum statistical mechanics, Nauka, Moscow, \mbox{384 p.} (1984). %
[N.N.\,Bogolyubov and N.N.\,Bogolyubov, Jr., Introduction to quantum statistical mechanics, 2nd ed., World Scientific,\,440 p.(2009)]. %
\bibitem{YuM07}
Yu.M.\,Poluektov, On the quantum-field description of many-particle Bose systems with spontaneously broken symmetry, Ukr. J. Phys. \textbf{52}, 578-594 (2007) [arXiv:1306.2103]. %
\bibitem{Bijl40}
A. Bijl,  The lowest wave function of the symmetrical many particles system, Physica \textbf{7}, 869-886 (1940). %
\bibitem{GirArn59}
M.\,Girardeau, R.\,Arnowitt, Theory of many-boson system: pair theory, Phys. Rev. \textbf{113}, 755-761 (1959). %
\bibitem{Went60}
G.\,Wentzel, Thermodynamically equivalent Hamiltonian for some many-body problems, Phys. Rev. \textbf{120}, 1572-1575 (1960). %
\bibitem{Lub62}
M.\,Luban, Statistical mechanics of a non-ideal boson gas: pair Hamiltonian model, Phys. Rev. \textbf{128}, 965-987 (1962). %
\bibitem{Tolm60}
V.V.\,Tolmachev, Temperature elementary excitations in a non-ideal Bose-Einstein system (in Russian), DAN SSSR vol.\textbf{135}(4), 825-828 (1960). %
\bibitem{YuM02}
Yu.M.\,Poluektov, Self-consistent field model for spatially inhomogeneous Bose systems, Low Temp. Phys. \textbf{28}, N6, 429p. (2002). %
\bibitem{Rubalko}
A. Rybalko, S. Rubets, E. Rudavskii, V. Tikhiy, S. Tarapov, R. Golovashchenko, and V. Derkach, Resonance \mbox{absorption} of microwaves in HeII: Evidence for roton emission. Phys. Rev. B. \textbf{76}, 140503(R) (2007). %
\bibitem{YuM14}
Yu.M.\,Poluektov, Absorption of electromagnetic field energy by the superfluid system of atoms with a dipole moment, Low Temp. Phys. \textbf{40}, N5, 389 p. (2014). %
\end{thebibliography}
\end{document}